\documentclass[10pt,twocolumn]{article}

\usepackage[
    letterpaper,
    top=0.75in,
    bottom=0.85in,
    left=0.75in,
    right=0.75in,
    columnsep=0.25in
]{geometry}

\usepackage[utf8]{inputenc}
\usepackage[T1]{fontenc}
\usepackage{microtype}

\usepackage{graphicx}
\usepackage{amsmath}
\usepackage{amssymb}
\usepackage{booktabs}
\usepackage{array}
\usepackage{tabularx}
\usepackage[table]{xcolor}
\usepackage{adjustbox}
\usepackage{pdflscape}
\usepackage{caption}
\usepackage{subcaption}
\usepackage{enumitem}
\usepackage{cite}

\usepackage{stfloats}
\usepackage{balance}

\usepackage[
    colorlinks=true,
    urlcolor=blue,
    linkcolor=black,
    citecolor=black
]{hyperref}
\setlist{
    leftmargin=*,
    itemsep=2pt,
    topsep=3pt,
    parsep=0pt
}

\newcommand{\Y}{\cellcolor{green!18}\textbf{Y}}
\newcommand{\Ptl}{\cellcolor{yellow!22}\textbf{P}}
\newcommand{\N}{\cellcolor{gray!12}\textbf{N}}

\newcolumntype{L}[1]{>{\raggedright\arraybackslash}p{#1}}
\newcolumntype{C}[1]{>{\centering\arraybackslash}p{#1}}

\usepackage{authblk}

\usepackage[ruled,vlined,linesnumbered]{algorithm2e}

\title{\vspace{-0.45in}Design-Intent Compilation for Heterogeneous Fabrication}

\author[1,3]{Charles Wade}
\author[2]{Devon Beck}
\author[1]{Robert MacCurdy\thanks{Corresponding Author: maccurdy@colorado.edu}}

\affil[1]{University of Colorado Boulder, Boulder, Colorado, USA}
\affil[2]{The Charles Stark Draper Laboratory, Inc., Cambridge, Massachusetts, USA}
\affil[3]{Draper Scholars, The Charles Stark Draper Laboratory, Inc., Cambridge, Massachusetts, USA}

\date{}

\begin{document}

\twocolumn[
\begin{@twocolumnfalse}
\maketitle
\begin{abstract}
Computational fabrication workflows increasingly require designers to specify spatially varying behavior, appearance, material composition, and process state. Yet most workflows still force these intentions into printer-specific representations, such as material fractions, voxel labels, mesh partitions, or slicer settings. This coupling makes heterogeneous designs difficult to reuse across fabrication systems because each backend requires a different realization of the same intended property. We present a compiler architecture for heterogeneous fabrication that treats computational fabrication design as a staged, typed lowering problem. A source design consists of implicit geometry and named, typed spatial attribute fields over a shared object domain. These attributes may encode measured data, visual appearance, target mechanical behavior, material recipes, process parameters, or other user-specified attributes. Translation models then derive compiler-required realization attributes from available source or intent attributes, and backend compilers emit machine-facing outputs such as voxel material stacks, process-annotated G-code, or configured slicer project files. This organization separates source modeling, attribute translation, and backend compilation, allowing a design to remain expressed in fabrication-agnostic semantic terms while each fabrication toolchain determines how those terms are translated into executable instructions. We validate the method through fabricated objects and real-world measurements across sampled volumetric data, CT-derived visual and mechanical models, Shore-hardness fields, and full-color fields, and by automatically implementing these desired attributes via material jetting and material extrusion. These results show that one source design can be lowered into distinct material, process, and slicer representations without rewriting the design for each target. We provide the representation, translation framework, compiler interface, and demonstrated workflows as an open-source Python package to accelerate research in heterogeneous computational fabrication.
\end{abstract}

\vspace{1em}

\end{@twocolumnfalse}
]

\section{Introduction}

Heterogeneous additive manufacturing (AM), a strategy to enable functional-as-printed objects, requires designers to specify more than shape. Depending on the application, a printed object may need spatially varying appearance, mechanical behavior, material composition, or process state. Yet many fabrication workflows still ask designers to express these intentions in printer-specific terms: material fractions, voxel labels, mesh partitions, temperatures, or slicer settings. These quantities are necessary for fabrication, but they are poor primary representations of design intent. A printer or print preparation software may require material mixtures, discrete voxel assignments, or process-control commands to realize a design. However, a designer may want to specify a target mechanical property (e.g., stiffness or elasticity), preserve color across fabrication processes, or derive printable properties from CT radiodensity data. When the designer must author printer-specific quantities directly, the design becomes coupled to a specific fabrication backend and the intended physical property is obscured.

This coupling of fabrication process to intended properties constrains designers by requiring them to express intent in process-specific terms, a limitation further complicated by the fact that different fabrication systems may realize the same semantic attribute through different mechanisms. For example, Shore hardness can be realized through calibrated rigid--soft material fractions on material jetting systems or through temperature-controlled foaming on material extrusion systems. Similarly, color can be realized as voxel-scale material assignments in inkjet printing or as halftoned color regions in filament-based workflows. In each case, the printer-specific attributes differ, but the source attribute specification can remain unchanged. This observation motivates the central argument of this paper: heterogeneous fabrication should treat material mixtures and process parameters as derived attributes, not as the primary design representation.

Prior work has made heterogeneous objects increasingly representable and printable. Functional and volumetric representations describe spatially varying fields, while programmable fabrication methods produce material, voxel, or process-specific outputs~\cite{pasko_constructive_2001,pasko_function_1995,vidimce_openfab_2013,vidimce_foundry_2016,chen_spec2fab_2013,wade_openvcad_2024}. However, many existing workflows bind the design to a particular attribute type, material model, or printing system. For example, Stratasys Digital Materials workflows allow designers to request Shore hardness values and automatically compute soft--rigid material mixtures, but only within a proprietary PolyJet preparation pipeline and only for spatially uniform values. The same intent cannot be directly retargeted to another process that realizes hardness through a different physical mechanism. A more general representation should let designers specify object intent independently from the printer-specific attributes used to realize it.

We frame heterogeneous fabrication as a staged translation problem inspired by compiler-based workflows. In conventional computing, a programmer writes source code in a high-level language that expresses what a program should compute; compilers then lower that program through target-dependent representations into instructions that execute on a specific computer architecture. The programmer does not normally write machine instructions directly. Instead, the language, compiler, libraries, and architecture together determine how high-level intent becomes executable behavior. Our work applies this separation of concerns to fabrication. 

This paper presents a field-based representation and compilation workflow that implements this view. A design consists of implicit geometry and named, typed spatial attribute fields over a shared object domain. A designer may directly specify attributes at either a high level, such as hardness, modulus, toughness, color, or radiodensity, or a lower level, such as material volume fractions, temperatures, flow rates, or slicer settings. Our method defines how these fields compose under modeling operations, how conflicts and undefined regions are resolved, and how derived attributes are computed from existing fields. Translation models encode calibrated, empirical, analytical, or user-defined mappings between attribute types. The resolver uses these models to derive compiler-required attributes when the source design does not provide them directly.

We validate the method through fabricated examples that test whether requested attributes survive translation and compilation. The experiments use a range of source inputs and target outputs, including analytic fields, sampled volumetric data, material-jetting outputs, color fabrication workflows, and material-extrusion process control. In several cases, we compile the same source design to distinct fabrication domains without rewriting the design in printer-specific terms. These examples show that intent-level attributes can be preserved through backend-specific realization, while also exposing the different constraints, fidelities, and artifacts of each fabrication process.

The contribution of this work is the staged translation approach to heterogeneous fabrication and the representation, resolver, and compiler interfaces that operationalize our approach. Rather than asking designers to author printer-specific material recipes or process schedules, the proposed workflow preserves intent as typed spatial fields and derives realization attributes only when a target compiler requires them. This enables design-intent-preserving cross-compilation across material jetting and material extrusion (MEX) workflows, while keeping the implementation extensible to new attribute types, translation models, and fabrication compilers. We provide an open-source Python implementation so that new attribute types, translation models, and fabrication compilers can be added by the research community.

\section{Related Work}

Heterogeneous additive manufacturing requires representations that describe more than boundary shape. A fabrication-ready design may contain spatially varying attributes such as material composition, color, opacity, mechanical behavior, sampled image data, process state, or slicer-level control. Prior work has developed many of these capabilities, but often under different assumptions about the properties that the designer can author and what the fabrication system can consume. Some methods emphasize expressive heterogeneous fields; others emphasize practical translation to a particular printer, material model, or voxel format. This section reviews these lines of work through the lens of the proposed representation: typed spatial attributes that can be composed, translated, and compiled to target-specific fabrication outputs.

\subsection{Heterogeneous Field Representations}

Early heterogeneous object models, summarized as M1 in Table~\ref{tab:heterogeneous-attribute-comparison}, establish that material and property information should be represented as part of the object, not only assigned after geometry creation. Kumar and Dutta introduce a materially graded object model in which material composition functions are attached to geometric regions and represented locally as volume fractions~\cite{kumar_solid_1997}. Kumar et al. generalize this view into an attribute-based object modeling framework in which geometry is the base attribute and other attributes, including material composition, microstructure, properties, and physical parameters, are defined over the object domain~\cite{kumar_framework_1999}. These works provide an important precedent for treating heterogeneous material information as first-class modeling data. However, their fabrication-facing representation remains centered on prescribed material allocation rather than a general mechanism for translating intent-level attributes into different material systems or process parameters.

Functional and hypervolume representations, summarized as M2, are foundational implicit modeling approaches. Pasko et al. introduce functional representations for geometry, where solids are described by mathematical fields rather than boundary meshes~\cite{pasko_function_1995}. HyperFun and constructive hypervolume modeling extend this idea to multidimensional attributes, allowing objects to carry fields such as density, opacity, color, or material-like quantities within a constructive modeling tree~\cite{pasko_hyperfun_1999,pasko_constructive_2001,hutchison_constructive_2008}. Our work builds on these concepts of implicit modeling and heterogeneous representations. 
These prior works focus on rendering and theoretical heterogeneous modeling. They begin to address the challenges of designing geometry and attributes but leave a gap in translating this design to fabrication, which we address with our method.

Other heterogeneous modeling methods make spatial material variation more accessible for CAD and manufacturing workflows. Source-based heterogeneous solid modeling, summarized as M3, defines material gradation from editable point, axial, planar, and surface sources, making graded distributions easier to specify relative to geometric features~\cite{siu_modeling_2002,siu_source-based_2002}. Trivariate spline and subdivision-volume methods, summarized as M4, represent geometry and volumetric attributes with continuous spline or subdivision fields and develop extraction or slicing procedures for additive manufacturing~\cite{martin_representation_2001,sasaki_adaptive_2017,luu_efficient_2019}. GraMMaCAD, summarized as M5, provides an interactive BRep-based interface for assigning spatially varying FGM distributions to imported CAD models~\cite{luu_grammacad_2022}. These systems provide important mechanisms for authoring, representing, and extracting heterogeneous material fields. However, the authored field is generally treated as the fabrication-relevant quantity itself: a material gradation, volumetric attribute, or FGM distribution to be sliced, sampled, or assigned. As a result, the representation does not clearly separate design-intent from fabrication-specific attributes, nor does it provide a general mechanism for deriving backend-specific material recipes or process parameters from a single fabrication-agnostic high-level source design. Our work addresses this gap by treating material fractions, colors, mechanical properties, sampled measurements, and process states as typed attributes that can be composed and translated according to compiler requirements.

Programmable multi-material fabrication methods, summarized as M6, move heterogeneous modeling closer to manufacturable output. OpenFab and Foundry introduce procedural volumetric material design workflows that generate printable multi-material representations~\cite{vidimce_openfab_2013,vidimce_foundry_2016}. OpenFab frames fabrication as a programmable pipeline and shows that procedural 3D textures and volumetric material descriptions can be transformed into material assignments for multi-material printing. In their workflow, however, the designer generally encodes the relationship between object behavior and material assignment directly in the design. OpenFab provides a powerful way to author heterogeneous material distributions, but it does not separate intent-level attributes from machine- or material-specific derived attributes.

Spec2Fab, summarized as M7, addresses the complementary problem of translating user-facing specifications into printable material assignments through a reducer--tuner framework~\cite{chen_spec2fab_2013}. It is one of the closest conceptual predecessors to our work because it recognizes that designers often want to specify perceptual or functional goals rather than low-level printer materials. Its translation, however, is organized around a comparatively monolithic optimization pipeline that couples the specification, material search, and target fabrication process to inkjet material assignment. Our work adopts the design-by-intent motivation, but represents translation models as modular field operators over typed attributes. This allows an authored field, such as shore hardness, modulus, toughness, RGBA color, or radiodensity, to be resolved differently depending on the selected compiler.

Voxel-based fabrication frameworks, summarized as M8, provide a direct route to manufacturable heterogeneous objects. VoxelFuse stores geometry and material information in voxel grids and supports multi-material design, manufacturing constraints, and simulation-oriented workflows~\cite{brauer_automated_2020}. Voxels naturally encode occupancy and material labels, and many printers and simulation tools can consume sampled volumetric data. The limitations are that sampling fixes the resolution early and that voxel-based representations can degrade under affine transformations. High-resolution material jetting can require very large voxel grids, while toolpath-based processes still require additional translation from voxel labels into slices and G-code. Voxel methods therefore provide practical sampled representations, but do not by themselves preserve resolution-independent design intent or support retargeting one high-level source design to multiple realization fields.

OpenVCAD, summarized as M9, combines signed-distance geometry with implicit material volume-fraction fields for multi-material additive manufacturing~\cite{wade_openvcad_2024}. Later extensions add a Python interface, simulation integration, lattice workflows, slicer-project export, and gradient-informed slicing for material extrusion~\cite{wade_implicit_2025,wade_implicit_2025-1,wade_slicer_project_control}. These works show the value of implicit geometry and implicit material fields for resolution-independent heterogeneous fabrication. Their primary heterogeneous representation, however, remains material volume fractions. Some process controls are supported, but they are introduced through a representation originally built around material fractions. Our  work makes the distinction explicit: material fractions are one possible derived attribute, not a universal design representation. Typed attributes allow properties, measurements, visual fields, material recipes, and process states to coexist and be translated according to manufacturing requirements.

Table~\ref{tab:heterogeneous-attribute-comparison} summarizes these methods. The comparison is not intended to rank generality in the abstract, but to identify which capabilities were demonstrated in the cited works. Prior methods support important subsets of heterogeneous modeling and fabrication. We are not aware of a prior system that jointly demonstrates typed implicit attributes, attribute conflict semantics, attribute derivation, process attributes, and cross-fabrication compilation from a single fabrication-agnostic source design.

\begin{table*}[t]
\centering
\caption{
Comparison of representative heterogeneous modeling and fabrication methods. Ratings indicate demonstrated capabilities in the cited works; they are not intended to rank generality in the abstract or imply that a feature was a goal of that work. Method groups are M1: cellular and attribute object models~\cite{kumar_solid_1997,kumar_framework_1999}; M2: FRep, HyperFun, and constructive hypervolume models~\cite{pasko_function_1995,pasko_hyperfun_1999,pasko_constructive_2001,hutchison_constructive_2008}; M3: source-based heterogeneous solid modeling~\cite{siu_modeling_2002,siu_source-based_2002}; M4: trivariate spline and subdivision-volume models~\cite{martin_representation_2001,sasaki_adaptive_2017,luu_efficient_2019}; M5: GraMMaCAD~\cite{luu_grammacad_2022}; M6: OpenFab and Foundry~\cite{vidimce_openfab_2013,vidimce_foundry_2016}; M7: Spec2Fab~\cite{chen_spec2fab_2013}; M8: VoxelFuse~\cite{brauer_automated_2020}; and M9: OpenVCAD~\cite{wade_openvcad_2024,wade_implicit_2025,wade_implicit_2025-1,wade_slicer_project_control}. For I1, `Y' indicates that a usable implementation is currently obtainable through public download or by request. The comparison highlights that prior methods support important subsets of heterogeneous modeling, but do not jointly demonstrate typed implicit attributes, attribute conflict semantics, attribute derivation, process attributes, and cross-fabrication compilation.
}
\scriptsize
\setlength{\tabcolsep}{3.2pt}
\renewcommand{\arraystretch}{1.18}
\begin{adjustbox}{max width=\linewidth}
\begin{tabular}{L{0.30\linewidth} *{10}{C{0.048\linewidth}}}
\toprule
\textbf{Feature} &
\textbf{M1} &
\textbf{M2} &
\textbf{M3} &
\textbf{M4} &
\textbf{M5} &
\textbf{M6} &
\textbf{M7} &
\textbf{M8} &
\textbf{M9} &
\textbf{This Work} \\
\midrule

\multicolumn{11}{>{}l}{\textbf{Representation features}} \\
\midrule

R1: Implicit geometry &
\N & \Y & \N & \N & \N & \Y & \N & \Ptl & \Y & \Y \\

R2: Explicit geometry / BRep &
\Y & \Y & \Y & \Y & \Y & \Y & \Y & \Y & \Y & \Y \\

R3: Sampled volume data &
\N & \Y & \N & \Y & \N & \N & \N & \Y & \Y & \Y \\

R4: Implicit attributes &
\Y & \Y & \Y & \Y & \Y & \Y & \Y & \N & \Y & \Y \\

R5: Sampled attributes &
\N & \Ptl & \N & \Y & \N & \Y & \Y & \Y & \Ptl & \Y \\

R6: Object / region composition &
\Y & \Y & \Y & \Y & \Y & \Y & \Y & \Y & \Y & \Y \\

R7: Attribute conflict rules &
\Y & \Y & \Y & \N & \Y & \Y & \Y & \Y & \Ptl & \Y \\

R8: Attribute derivation &
\Y & \Y & \N & \Ptl & \N & \Y & \Y & \N & \N & \Y \\

\midrule
\multicolumn{11}{>{}l}{\textbf{Attribute-type features}} \\
\midrule

A1: Rendering attributes &
\N & \Y & \Y & \Y & \Y & \Y & \Y & \N & \Y & \Y \\

A2: Physical attributes &
\Y & \Y & \N & \Y & \N & \Y & \Y & \N & \N & \Y \\

A3: Material attributes &
\Y & \Y & \Y & \Y & \Y & \Y & \Y & \Y & \Y & \Y \\

A4: Process attributes &
\N & \N & \N & \N & \N & \N & \N & \N & \Ptl & \Y \\

\midrule
\multicolumn{11}{>{}l}{\textbf{Output-modality features}} \\
\midrule

O1: Rendering &
\N & \Y & \Y & \Y & \Y & \Y & \Y & \Y & \Y & \Y \\

O2: Inkjet / PolyJet: color &
\N & \N & \N & \Y & \N & \Y & \Y & \N & \N & \Y \\

O3: Inkjet / PolyJet: material allocation &
\N & \N & \N & \Y & \Y & \Y & \Y & \N & \Y & \Y \\

O4: Material extrusion material allocation &
\Y & \N & \N & \N & \Y & \N & \N & \Y & \Y & \Y \\

O5: Process parameters &
\N & \N & \N & \N & \N & \N & \N & \N & \Ptl & \Y \\

O6: Slicer settings &
\Ptl & \N & \Ptl & \N & \N & \N & \N & \N & \Ptl & \Y \\

O7: Cross-fabrication &
\N & \N & \N & \N & \N & \N & \N & \N & \N & \Y \\

\midrule
\multicolumn{11}{>{}l}{\textbf{Implementation availability}} \\
\midrule

I1: Implementation available for use &
\N & \Y & \N & \N & \Y & \N & \Y & \Y & \Y & \Y \\

\bottomrule
\end{tabular}
\end{adjustbox}
\label{tab:heterogeneous-attribute-comparison}
\end{table*}

\subsection{Programmatic and Data-Driven Authoring}

Many heterogeneous and procedural design methods use programming interfaces because spatial fields are difficult to author through direct manipulation alone. HyperFun provided a language for functional representation modeling~\cite{pasko_hyperfun_1999}. OpenFab used a shader-like language for procedural material behavior~\cite{vidimce_openfab_2013}. OpenSCAD popularized script-based constructive solid geometry for conventional geometric CAD~\cite{kintel_openscad_2010}. Early OpenVCAD versions likewise used a custom scripting language for implicit geometry and volume fractions~\cite{wade_openvcad_2024}, while later work moved the workflow into Python~\cite{wade_implicit_2025}. These methods demonstrate that code is an expressive design representation. However, a programming interface does not by itself define the semantic structure of heterogeneous fabrication. There remains a research gap in what properties are being authored in these code-based languages and in how code is translated into manufacturable formats. Extending these prior methods, our work demonstrates the design of arbitrary high-level attributes, intermediate translation between attributes, and cross-compilation of a single design into multiple manufacturing-process specific formats.

\subsection{Translation Models and Process Attributes}

A design-intent workflow also requires models that translate desired properties into realizable material or process fields. One source for these models is recent material-calibration work that provides reusable local mappings from target properties to printable recipes. Smith et al., for example, introduce a data-driven mapping from mechanical targets such as modulus and toughness to digital material mixtures~\cite{smith_digital_2024}. Such models are useful in our setting because they can be treated as field operators: if the source design contains spatially varying mechanical attributes, the same calibrated relationship can be evaluated throughout the object to produce compiler-required material fractions.

Process parameters further expand the set of attributes that a compiler may require and are therefore useful in translation models. Spatially varying behavior can arise from material assignment, but also from laser power, extrusion rate, print speed, beam focus, nozzle temperature, material mixing ratio, or flow rate, depending on the printing modality~\cite{loh_overview_2018,wade_implicit_2025-1,damanpack_porous_2021,tammaro_microfoamed_2022,lalegani_dezaki_soft_2023}. Mixing systems in material extrusion and direct ink writing likewise demonstrate that material ratio can be a controllable process state rather than only a static material label~\cite{ortega_active_2019,liao_active_2024,pelz_multi-material_2021,li_fabricating_2018}. These examples motivate process attributes as first-class fields in heterogeneous design. A compiler may require temperature, flow rate, mixture ratio, or slicer settings rather than a material-fraction vector, and a staged translation workflow can derive these realization attributes from higher-level intent when a valid model exists.

\subsection{Voxel, Color, and Medical Image Fabrication}

Material jetting and PolyJet-style systems motivate sampled output because they can assign discrete materials at droplet or voxel scale. Color 3D printing has therefore developed sophisticated printer-dependent reproduction methods, including scattering-aware texture reproduction, geometry-aware compensation, neural acceleration of scattering prediction, and gradient-based material-assignment optimization~\cite{elek_scattering-aware_2017,sumin_geometry-aware_2019,rittig_neural_2021,nindel_gradient-based_2021}. These methods address the final conversion from desired appearance to printer-specific material arrangements. In our workflow, color reproduction is treated as a compiler-side realization problem: RGBA color is one typed attribute that may be authored, sampled, or derived, and specialized color pipelines can then convert that attribute into process-specific voxel or slicer outputs.

Medical image-based fabrication presents a related sampled-data problem. Patient-specific models are used for pre-surgical planning, procedural rehearsal, education, and clinical communication~\cite{valls-esteve_exploring_2023,zoabi_3d_2022,gonzalez-lopez_integration_2024}. Many workflows convert CT or MRI data into segmented surfaces using marching cubes or related surface extraction methods~\cite{bucking_medical_2017,kamio_dicom_2020,fogarasi_algorithms_2022,lorensen_marching_1987,newman_survey_2006,mandaliana_3d_2019,martel_voxel2mesh_2020}. These workflows are practical, but they reduce volumetric image data to boundaries or labels before fabrication. Voxel-based medical printing preserves more of the original scan information and has been used to map medical images to full-color material-jetting outputs~\cite{hosny_improved_2018,jacobson_defining_2022,jacobson_inner_2023,jacobson_hybrid_2023,jacobson_voxel_2022}. Our work follows the volumetric view, but treats scan intensity as a typed sampled source attribute rather than as a fixed route to one printer color space. The same input radiodensity field can be mapped to outputs such as visual attributes, mechanical attributes, or material recipes depending on the selected translation models and compiler.

\subsection{Summary}

Prior work provides strong components for heterogeneous fabrication: field-based representations, procedural material design, voxel fabrication, specification-to-material optimization, process-aware slicing, color reproduction, and medical image printing. The remaining gap is a representation and compilation model that makes these capabilities composable at the level of design intent. We address this gap by framing heterogeneous design as a staged compilation process in which high-level attribute specifications are lowered through interchangeable fabrication toolchains. This organization allows the same high-level design intent to be retargeted across fabrication modalities without rewriting the source design in the vocabulary of each machine.

\section{Methods}

We represent heterogeneous objects as spatial fields over a common domain. A source design consists of implicit geometry and a set of named, typed attribute fields. Later stages operate on this same representation: modeling operations compose fields, translation models derive additional fields when needed, and fabrication compilers sample the resolved fields into process-specific outputs. We first define the source design, since composition, translation, and compilation all depend on this mathematical framework.

\subsection{A Field-Based Representation for Heterogeneous Objects}
\label{sec:field_based_representation}

Let an object be defined by a scalar implicit geometric field
\begin{equation}
    \phi : \mathbb{R}^3 \rightarrow \mathbb{R},
\end{equation}
where \(p=(x,y,z)\in\mathbb{R}^3\) is a three-dimensional spatial point and \(\phi(p)\in\mathbb{R}\) is its signed distance to the object's boundary. We use the convention that $\phi(p) < 0$ denotes points inside the object, $\phi(p) = 0$ denotes points on the boundary, and $\phi(p) > 0$ denotes points outside the object. The solid domain is therefore
\begin{equation}
    \Omega = \{p \in \mathbb{R}^3 \mid \phi(p) \leq 0\}.
\end{equation}
This convention fixes the geometric meaning of all subsequent field evaluations. Objects may be authored directly from analytic signed distance functions, but the representation does not require analytic primitives. Triangular meshes and CAD boundary representations can also enter the workflow by first converting them to signed distance fields, preserving a single spatial query interface for downstream heterogeneous modeling.

A source design pairs this geometric field with a finite set of named, typed attribute fields,
\begin{equation}
    \mathcal{A} = \{(n_i, T_i, A_i)\}_{i=1}^{m},
\end{equation}
where \(n_i\) is a semantic name, \(T_i\) is the return type, and
\begin{equation}
    A_i : \Omega \rightarrow T_i
\end{equation}
is an attribute field that returns a value at each point inside or on the object. An attribute may be implemented by an evaluator
\begin{equation}
    \widetilde{A}_i : \Omega \times \mathbb{R} \rightarrow T_i,
\end{equation}
such that
\begin{equation}
    A_i(p) = \widetilde{A}_i\bigl(p,\phi(p)\bigr).
\end{equation}
Here, the second argument is a one-dimensional scalar containing the signed distance from \(p\) to the object's boundary. This formulation allows attribute values to depend both on spatial position and on distance from the boundary while preserving \(A_i\) as a field defined over the solid domain. The name identifies the field's meaning, such as color, hardness, radiodensity, material composition, or process state. The type specifies the structure of the returned value. In this work, we demonstrate scalar attributes, three- and four-dimensional vector attributes, and material-fraction vectors, although the representation is not restricted to these types.

\begin{figure}[th!]
    \centering
    \includegraphics[width=\linewidth]{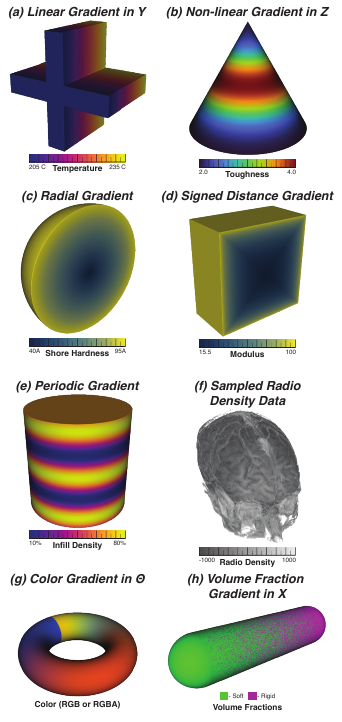}
    \caption{
    Representative attribute fields expressible through the proposed field representation. Analytic attributes can vary in Cartesian coordinates (a), (b), and (h); spherical coordinates (c); cylindrical coordinates (g); signed distance from the boundary (d); or periodic functions (e). Sampled attributes, such as CT-derived radiodensity (f), enter the same framework through interpolated field evaluators. The volume-fraction example in (h) is visualized as discrete material assignments for clarity, but the underlying attribute is a continuous material-fraction field. These examples illustrate possible evaluators rather than an exhaustive set of supported attributes.
    }
    \label{fig:attribute_showcase}
\end{figure}

Attribute evaluators may be analytic, sampled, or derived. Analytic attributes are functions of position, object parameters, or geometric quantities. As shown in Figure~\ref{fig:attribute_showcase}, they can express parametric behavior such as Cartesian, radial, cylindrical, periodic, or boundary-dependent gradients. Sampled attributes incorporate externally generated volumetric data, such as CT or MRI volumes, image stacks, OpenVDB grids, or simulation outputs. Sampled data can be treated the same way as analytic data after interpolation defines an evaluator over $\Omega$: it can be queried, composed, translated, and compiled. Derived attributes are produced by translation models described in Section~\ref{sec:attribute_translation}.

This evaluator model is the representation-level abstraction. A source design is not a mesh, voxel stack, material partition, or printer program. It is a queryable field object whose geometry and heterogeneous properties share a common spatial interface, while permitting discontinuities and definitions over multiple disjoint regions. This abstraction lets analytic design intent and sampled data coexist in one object, defers printer-specific material and process decisions until later stages, and provides the mechanism by which composition, translation, and compilation operate.

\subsection{Composition Semantics for Attribute Fields}
\label{sec:composing_attribute_fields}

The field representation in Section~\ref{sec:field_based_representation} defines how a single heterogeneous object is evaluated. Design workflows, however, build objects compositionally. Designers define primitives, transform them into position, and combine them into larger structures. We represent this process as a design graph whose leaf nodes define geometric primitives, and whose interior nodes transform or combine the fields returned by their children. Evaluation at a point proceeds recursively through the graph: each node queries its children and applies a local operator to the returned geometry and attributes.

Transforms act by changing the coordinate frame in which a child is queried. Let $T$ be a rigid transform or uniform scale, let $s$ be the corresponding scale factor, and let $c$ denote the child node. The transformed geometry is evaluated as
\begin{equation}
    \phi_T(p) = s\,\phi_c(T^{-1}p),
\end{equation}
with $s=1$ for rigid transforms. Attributes follow the same inverse-coordinate rule:
\begin{equation}
    A_{i,T}(p,\phi_T(p)) =
    A_{i,c}(T^{-1}p,\phi_c(T^{-1}p)).
\end{equation}
Thus, transforms do not resample or rewrite the design. They only change the spatial query passed to the child evaluator. We restrict this rule to rigid transforms and uniform scaling because these operations preserve the signed-distance.

Boolean composition combines the fields returned by multiple children. For geometry, we use standard signed-distance Boolean operations for union, intersection, and difference~\cite{pasko_function_1995}. These operations define the composed solid domain for the node. The additional question for heterogeneous modeling is not how to combine geometry, but how attributes should behave when geometric operations combine regions that carry different fields. We handle this with two local semantic rules: conflict resolution, which applies when several children define the same attribute at the same point, and attribute availability, which records when a queried attribute is absent from part of the composed object, as discussed in the next section.

\subsubsection{Hierarchy, Overrides, and Attribute Conflicts}
\label{sec:attribute_conflicts}

Attributes can be attached at any node in the design graph. A parent-level attribute may introduce a new named field over the composed geometry, or override a field with the same name and type provided by its children. Parent override is the simplest hierarchical rule: if a node defines attribute $a$ over its output domain, that field is exposed for $a$ independent of child values. This lets a designer construct geometry from several components while assigning one attribute field over the result.

\begin{figure}[th!]
    \centering
    \includegraphics[width=\linewidth]{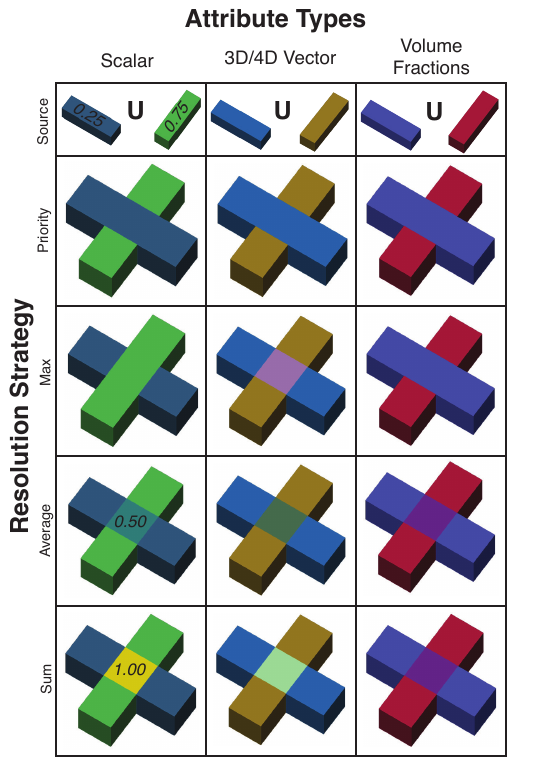}
    \caption{
    Attribute conflict resolution for overlapping child nodes. The source row shows two independently authored objects before union. Subsequent rows show the same geometric union resolved using priority, maximum, average, and a custom summation operation. Columns show scalar, 3D/4D vector, and material volume-fraction attributes. Conflict resolution is defined per attribute type; if no resolver is specified, child-order priority is used. The geometry is unchanged across rows, while the resolved overlap value depends on the selected semantic rule.
    }
    \label{fig:attribute_composition_conflicts}
\end{figure}

If no parent override is supplied, Boolean composition can create attribute conflicts. A conflict occurs only when multiple contributing children define attributes with the same name and type at the same point. Attributes with different names or different types do not conflict; they coexist as separate fields. Figure~\ref{fig:attribute_composition_conflicts} shows this case for two overlapping bars whose union produces an overlap region with multiple candidate values.

We formulate conflict resolution as a typed reduction over candidate attribute samples. For an attribute identified by name and type, let
\begin{equation}
\label{eq:attribute_conflicts_set}
    \mathcal{V}_{a}(p)=\{v_1, v_2, \ldots, v_n\}
\end{equation}
be the values sampled from contributing children at point $p$. A resolver for attribute $a$ is a function
\begin{equation}
    v_{\mathrm{out}} = R_a(\mathcal{V}_{a}(p)),
\end{equation}
that maps the candidate samples to a single value of the same type. The resolver is invoked only when multiple contributing children define the same named and typed attribute at $p$.

One resolver is child-order priority: the first child, in the order added to the node, that defines the queried attribute at $p$ supplies the value. This rule is deterministic and valid for any attribute type. Other resolvers can be selected when they match the semantics of the field, including extrema for ordered scalar values, averaging for compatible scalar or vector fields, and custom user-defined operations. Material-fraction fields may use specialized resolvers when a mixture operation is intended. Any such resolver must return a valid material-fraction vector, preserving nonnegative components that sum to one.

Conflict rules define the meaning of heterogeneous composition. The same geometric union can yield different attribute fields depending on whether the overlap selects one child, preserves an extremum, averages compatible values, or applies a custom operation. By making these rules explicit and local to the design graph, the proposed representation gives attributes compositional semantics under the same hierarchy, transforms, and Boolean operations used to construct geometry.

\subsubsection{Attribute Availability and Undefined Regions}
\label{sec:attribute_availability}

Composition can also produce regions where a queried attribute is not defined. This is distinct from a conflict. Conflict resolution chooses among multiple values; availability records whether any value exists. Because attributes are attached to nodes in the design graph, an object can be geometrically valid while carrying different attribute sets in different regions. If $\mathcal{V}_{a}(p)$ (as defined in Eq.~\ref{eq:attribute_conflicts_set}) is empty, then attribute $a$ is undefined at $p$.

Undefined attributes are a normal intermediate state during modeling. They become errors only when a downstream translation model or compiler requires the missing attribute at a point where it is undefined. This distinction allows different parts of a source design to be authored at different semantic levels and made compatible later. A designer may provide a default field, introduce a parent-level override, or apply a translation model that derives the missing attribute before the design reaches a downstream consumer.

\begin{figure}[h!]
    \centering
    \includegraphics[width=\linewidth]{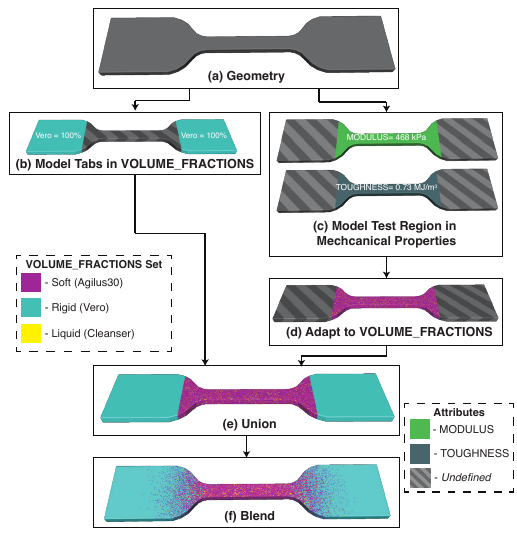}
    \caption{
    Region-specific attribute authoring in a single dog-bone design. The tabs are authored directly as rigid volume fractions, while the central gauge region is authored using target modulus and toughness. These choices intentionally leave some attributes undefined in some regions during authoring. Before compilation, the mechanical attributes in the gauge region are translated into volume fractions, making the regions compatible for union and blending. Undefined attributes are valid intermediate states unless they remain unresolved when required by a downstream translation or compiler.
    }
    \label{fig:hybrid_attribute_modeling}
\end{figure}

Figure~\ref{fig:hybrid_attribute_modeling} illustrates this workflow with a heterogeneous dog-bone specimen. The gripping tabs are specified directly as rigid volume fractions, while the central gauge region is specified by target modulus and toughness. During authoring, the tabs do not define modulus or toughness, and the central region does not yet define volume fractions. Before compilation, the pair of mechanical attributes in the gauge region is lowered into volume fractions by a translation model, as described in Section~\ref{sec:attribute_translation}. The tabs and gauge region then expose compatible material-fraction fields and can be combined into one material-recipe design.

Making availability explicit separates geometric validity from attribute completeness. A composed object may be valid as geometry even when some attributes are unavailable in some regions. Only the attributes required by a later translation or compiler must be defined over the domain consumed by that downstream operation.

\subsection{Lowering Attributes for Fabrication}
\label{sec:attribute_translation}

The preceding sections define how attributes are represented and composed. We now define how an authored attribute becomes useful to a particular fabrication process. The key distinction is between what an attribute means and how a backend realizes it. 

Attributes play three roles in the lowering workflow. \emph{Source attributes} describe measured, simulated, or externally generated data, such as CT-derived radiodensity or OpenVDB density and temperature grids. \emph{Design-intent attributes} describe the desired object property or behavior, such as Shore hardness, modulus, toughness, or density. \emph{Realization attributes} describe the material or process state required by a fabrication backend, such as volume fractions, temperature, flow rate, or slicer-level settings. These categories describe how an attribute is used in a workflow, not a fixed property of the name. RGBA color may be authored as visual intent, imported as source data, or consumed directly by a color-printing backend. Similarly, volume fractions often act as a realization attribute, but designers may also author them directly when fabricating calibration coupons used to measure material behavior and fit new translation models.

An attribute translation model is a field operator that derives one or more new named attributes from existing attributes. This differs from the conflict resolvers in Section~\ref{sec:attribute_conflicts}: resolvers reduce multiple candidate values of the same attribute, while translations create new attributes from different inputs. For example, a translation may lower shore hardness into volume fractions, map radiodensity into RGBA color, or compute temperature and flow rate from a target density field. The translation node preserves the child geometry and exposes the original attributes together with any translated outputs.

Formally, let a translation model $M_j$ define a local mapping from input attribute samples to output attribute samples. For an output attribute $\beta$, the translated evaluator is
\begin{equation}
\begin{aligned}
A_{\beta}(p,\phi(p))
&=
M_j\!\left(
    \{A_{\alpha_i}(p,\phi(p))\}_{i=1}^{k}
\right), \\
    &\hspace{2em} p \in \Omega,
\end{aligned}
\end{equation}
where $M_j$ is the analytical, empirical, calibrated, or user-defined translation model and ${A_{\alpha_1}, A_{\alpha_2}, \ldots, A_{\alpha_k}}$ denotes the set of input attributes. A model may produce multiple output attributes (e.g., $A_{\beta_1}, A_{\beta_2}, \ldots$) such that each output is evaluated as a field over the same spatial domain. This definition of translation models allows functional composition, mapping, or remapping of attributes that is propagated only when the output is queried.

Many fabrication translations encode inverse design relationships. The designer specifies a desired property field, and the translation computes the material recipe or process state expected to realize that property on a given fabrication system. For example, a calibrated material model can lower modulus and toughness into volume fractions for a multi-material inkjet workflow~\cite{smith_digital_2024}. Other translations connect source data to design intent before fabrication-specific lowering occurs. A CT-derived radiodensity field may be mapped directly to RGBA color for visual anatomical printing, or first mapped to mechanical attributes that are later lowered into material fractions as we will demonstrate in the results. In each case, translation models act as functional lowering rules between semantic fields.

\subsubsection{Attribute Resolver}
\label{sec:attribute_translation_graph}

Designers may insert translation nodes manually when the intended fabrication path is known. However, manual insertion can make a source design unnecessarily backend-specific. A design authored in terms of shore hardness should not need to contain separate hard-coded translation chains for every material system and printer it might later target. To avoid this coupling, we use an attribute resolver that selects and inserts translation nodes from a registered translation graph.

\begin{figure}[th!]
    \centering
    \includegraphics[width=\linewidth]{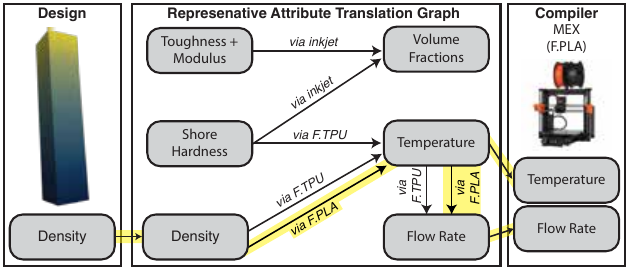}
    \caption{
    Attribute translation graph used to satisfy compiler-required attributes. The source design provides density, while the selected material-extrusion compiler for foaming PLA requires temperature and flow rate. Nodes represent typed attributes or paired attribute requirements, and directed edges represent registered translation models. Workflow tags, such as \texttt{via F.PLA}, \texttt{via F.TPU}, and \texttt{via inkjet}, restrict the resolver to translations compatible with the selected fabrication toolchain. The highlighted route shows the selected lowering path used to expose the attributes required by the compiler.
    }
    \label{fig:attribute_translation_graph}
\end{figure}

The translation graph is a dependency graph over named, typed attributes and registered translation models. It is distinct from the design graph: the design graph defines spatial composition, while the translation graph defines how attributes can be lowered into other attributes. As shown in Figure~\ref{fig:attribute_translation_graph}, nodes represent single attributes or paired attribute requirements, such as modulus+toughness. Directed edges represent translation models. Each edge records its required input attributes, output attributes, model constructor, and optional workflow tags. Workflow tags restrict a translation to a particular material system, calibration context, or backend workflow, such as \texttt{via F.PLA} (foaming PLA) versus \texttt{via F.TPU} (foaming TPU).

Algorithm~\ref{alg:attribute_resolver} shows this process. Given a source design and a target compiler, the resolver performs dependency resolution over the available attribute set. It begins with the attributes already exposed by the design and the attributes required by the compiler. A translation is eligible when all of its input attributes are available and its workflow tags are compatible with the requested workflow. Applying the translation adds its output attributes to the available set. The resolver continues until all required realization attributes are available, then wraps the original source design with the selected translation nodes. The result is not a new representation; it is the same queryable field object augmented with the operators needed to expose the requested realization attributes.

\begin{algorithm}[t]
\caption{Resolve compiler-required attributes}
\label{alg:attribute_resolver}
\KwIn{Source design $D$, available attributes $\mathcal{S}$, compiler-required attributes $\mathcal{R}$, registered translations $\mathcal{M}$, workflow tags $\mathcal{G}$}
\KwOut{Adapted design $D'$ exposing all attributes in $\mathcal{R}$}
$D' \leftarrow D$\;
$\mathcal{A} \leftarrow \mathcal{S}$\;
\While{$\mathcal{R} \nsubseteq \mathcal{A}$}{
    $\mathcal{E} \leftarrow \{M \in \mathcal{M} \mid \mathrm{inputs}(M) \subseteq \mathcal{A} \ \land\ \mathrm{tags}(M)\ \mathrm{match}\ \mathcal{G}\}$\;
    \If{$\mathcal{E} = \emptyset$}{
        \Return error: required attribute cannot be resolved\;
    }
    Select $M^\ast \in \mathcal{E}$ that advances an unresolved required dependency\;
    Wrap $D'$ with the translation node constructed by $M^\ast$\;
    $\mathcal{A} \leftarrow \mathcal{A} \cup \mathrm{outputs}(M^\ast)$\;
}
\Return $D'$\;
\end{algorithm}

This dependency-resolution step is the lowering stage of the proposed workflow. The source design may contain high-level intent or source data, while the compiler declares the realization attributes it needs. The resolver inserts only the translation models needed to satisfy those requirements. If no compatible translation path exists, the design cannot be compiled for that backend without additional authored attributes or registered models.

\subsection{Fabrication Compilation}
\label{sec:compiling_resolved_attribute_fields}

The preceding sections define how source designs are represented, composed, and lowered into realization attributes. Compilation is the final stage: a compiler consumes a resolved field object and emits the format required by a target fabrication or rendering workflow. In this work, compilers are not the primary novelty. We adapt established voxel, toolpath, and slicer-project export strategies into a common interface so that each compiler can declare the attributes it requires and sample the same resolved spatial representation.

In our proposed method, each compiler declares a set of required attributes:
\begin{equation}
    \mathcal{R}_b = \{\mathcal{A}_1,\mathcal{A}_2,\ldots,\mathcal{A}_q\},
\end{equation}
where $b$ denotes the target workflow. The resolver described in Section~\ref{sec:attribute_translation_graph} inserts translation nodes until the design exposes all attributes in $\mathcal{R}_b$. Then a compiler defines the mapping from the set of required attributes to workflow-specific output, denoted $O_b$:

\begin{equation}
    C_b : (\phi, \mathcal{R}_b) \rightarrow O_b,
\end{equation}
The compiler then samples geometry and attributes according to the target workflow: voxels for material jetting, toolpath locations for material extrusion, or sub-mesh regions for slicer-project control.

\subsubsection{Inkjet Compilation}
\label{sec:inkjet_compilation}

Voxel-based inkjet and material-jetting workflows require a discrete printable material assignment at each sampled voxel or droplet location. We adapt two inkjet compiler strategies into the attribute interface defined above. Both consume resolved volume fractions or RGBA color fields and convert continuous spatial attributes into material-specific voxel stacks.

The stochastic compiler, adapted from OpenVCAD~\cite{wade_openvcad_2024}, samples the resolved material distribution independently at each voxel and assigns one printable material according to that distribution. This approach is simple and readily parallelizable, but sampling each location independently can produce local statistical imbalances, resulting in visible clusters of one material despite a smoothly varying target composition~\cite{bacca_rodriguez_blue-noise_2008}. The structured compiler instead applies three-dimensional error diffusion inspired by the work of Brunton et al.~\cite{brunton_pushing_2015,brunton_cuttlefish_2023}. It assigns one material per voxel while propagating quantization error through the voxel grid, which improves local agreement between the requested continuous field and the accumulated printed material distribution. As shown in Table~\ref{tab:results_compilation_cost}, this improvement increases export time.

\subsubsection{Material Extrusion Compilation}
\label{sec:material_extrusion_compilation}

Material extrusion workflows require different compilation strategies because fabrication output is organized around paths, slicer regions, settings, and process-state changes rather than independent voxel assignments. We adapt two prior MEX compiler families into the proposed attribute interface: Gradient-Informed Toolpath Planning~\cite{wade_implicit_2025-1} and Slicer Project Control~\cite{wade_slicer_project_control}. Both consume resolved spatial attributes, but they target different levels of machine control.

Gradient-Informed Toolpath Planning compiles attributes directly into process-annotated G-code. The original method encoded process states, such as temperature, indirectly as material volume fractions~\cite{wade_implicit_2025-1}. We adapt the compiler to consume resolved temperature and flow rate fields directly during path generation, allowing spatial attributes to influence path construction and process scheduling without this intermediate encoding. This compiler is useful when the fabrication result depends on non-instantaneous process states, such as nozzle temperature for foaming filaments or mixture ratio for in-hotend filament mixing.

Slicer Project Control, proposed by Wade et al.~\cite{wade_slicer_project_control}, targets workflows where a conventional slicer performs final toolpath generation. The compiler discretizes resolved fields into sub-meshes or slicer-control regions and exports configured project files in which each region carries the appropriate slicer-level setting. This method supports attributes such as infill density, temperature, flow rate, and RGB color while preserving access to mature slicer functionality. We use it in the results to compare cross-compiled outputs produced from the same resolved design.

These MEX compilers therefore serve complementary roles. Gradient-informed slicing provides direct control over path generation and process-state scheduling, whereas Slicer Project Control uses slicer project files to route spatial attributes through existing slicing workflows. 

\subsection{Cross-Compilation Across Fabrication Toolchains}
\label{sec:cross_compilation}

Cross-compilation follows from the staged organization of the proposed workflow. Its value is that fabrication intent can be preserved independently of any particular machine or process, allowing designers to retarget, compare, or replace fabrication toolchains without rebuilding the design around each toolchain's process-specific controls. A design is authored once as geometry plus named, typed attributes. Translation models then lower those attributes into the material, process, or slicer-control fields required by a selected fabrication backend. Finally, compiler modules emit the machine-facing artifact for that backend. Unlike conventional workflows in which design intent is embedded directly in printer- or slicer-specific parameters, this structure organizes computational fabrication as a programming-oriented, multi-backend compilation workflow: designers author source-level fabrication intent, while target-specific compilers determine how that intent becomes printable instructions.

\begin{figure}[th!]
    \centering
    \includegraphics[width=\linewidth]{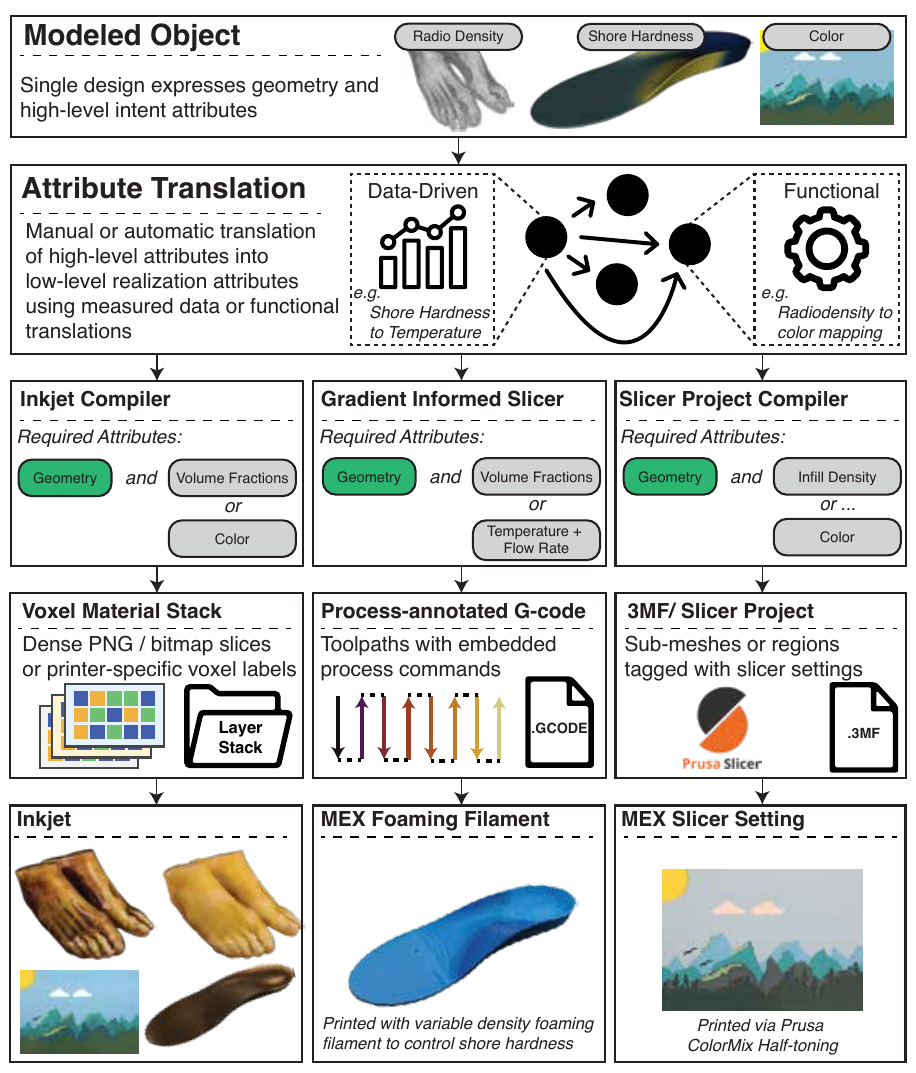}
    \caption{
    Overview of the proposed multi-backend compilation workflow for heterogeneous fabrication. A modeled object expresses geometry and high-level spatial attributes. Attribute translation lowers these fields into realization attributes. Backend compilers then request the attributes required by their target workflow and emit machine-facing outputs. Separating source modeling, attribute translation, and backend compilation enables the same design to be retargeted across fabrication systems.
    }
    \label{fig:compiler_contract}
\end{figure}

Figure~\ref{fig:compiler_contract} summarizes the complete method. The modeled object remains the source representation. Attribute translation provides the intermediate lowering layer, using either measured calibration data, inverse process models, or functional mappings. The compiler layer then consumes only the attributes required by the selected target and emits an output in the appropriate representation: voxel material stacks for inkjet printing, process-annotated G-code for gradient-informed material extrusion, or configured \texttt{.3MF} projects for slicer-based workflows. Because these stages remain separate, the same source design can be retargeted without rewriting the design in the vocabulary of each printer or slicer.

\section{Results}
\label{sec:results}

The results evaluate whether source designs authored in intent-level attributes can be lowered, compiled, and fabricated across distinct processes. We first specify the calibrated translation models used as reusable lowering rules in later examples. We then fabricate a set of diverse sampled volumetric attributes from simulation, medical, and industrial CT data using color inkjet compilation. These examples progress from single-process visual prints to a presurgical planning example that compiles the same sampled source data into both visual and mechanical material-jetting outputs. Finally, we use cross-compilation examples to test whether a single source design can be realized through different material systems and fabrication toolchains, first for Shore-hardness fields and then for full-color fields. Together, these results show how attribute-level specification can preserve design intent while exposing the different constraints and artifacts introduced by each compiler and process.

Detailed calibration procedures, fitted coefficients, lookup tables, and supporting validation data are provided in the supplementary information.

\subsection{Calibrated Attribute Translation Models}
\label{sec:results_translation_validation}

First, we describe the three translation models used to fabricate our results. Two are adopted from prior work, where they were developed for specific fabrication workflows rather than as components of a generalized attribute-modeling method. The third is a new inkjet material-jetting calibration introduced here. In our system, all three models serve as reusable translation layers that lower high-level physical-property fields into the material recipes or process parameters consumed by different compilers.

For mechanical-property specifications, we adapt the digital material model of Smith et al.~\cite{smith_digital_2024} as a translation from modulus and toughness to volume fractions. The output volume-fraction vector specifies mixtures of Vero, Agilus30, and the printer's non-curing cleaning fluid. Smith et al. established an inverse design mapping between target mechanical properties and printable material recipes, and showed that the reachable property range overlaps with soft-tissue mechanical properties. This translation is used later for the CT scan of feet example in Fig.~\ref{fig:results_feet}.

For material extrusion with foaming TPU, we adopt the calibrated process models of Wade et al.~\cite{wade_implicit_2025-1,wade_slicer_project_control}. This workflow uses two linked translations. The first maps target shore hardness to nozzle temperature. The second maps temperature to a compensatory flow rate. Together, these fields form the process-attribute pair consumed by the MEX compiler to control foaming behavior while maintaining geometric accuracy. In the cross-compilation results, this model pair provides the material-extrusion realization of the same Shore-hardness intent used for inkjet.

\begin{figure}[t]
    \centering
    \includegraphics[width=\linewidth]{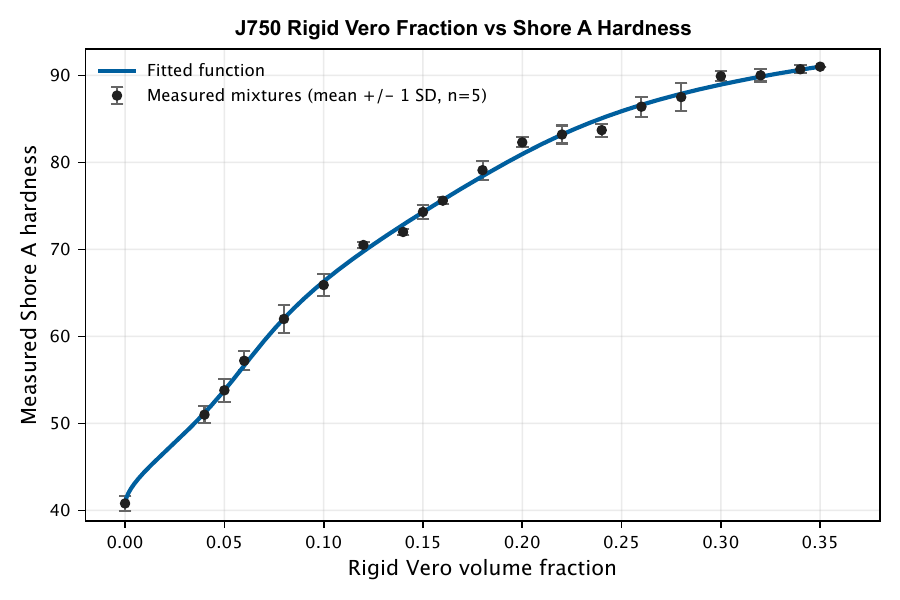}
    \caption{
    Inkjet Shore A hardness translation model for Vero/Agilus30 material jetting mixtures.
    Calibration specimens were printed with prescribed Vero volume fractions, with the remaining fraction assigned to Agilus30. Points show measured Shore A hardness values with error bars indicating one standard deviation across five specimens per mixture. The fitted inverse model defines the translation from target shore hardness to the volume fractions required by the inkjet material compiler.
    }
    \label{fig:results_cross_shore_hardness_j750_model_fit}
\end{figure}

We also introduce a new shore hardness to volume fractions model for inkjet with Vero and Agilus30. Calibration specimens were printed at prescribed Vero volume fractions from \(0\%\) to \(35\%\), with the remaining fraction assigned to Agilus30. Five specimens were printed for each mixture. The measured calibration range spanned \(40.8\) to \(91.0\) Shore A. Because the inkjet compiler requires material fractions, we fit the inverse relationship directly, yielding a translation from target shore hardness to the volume fractions consumed by the compiler. This model provides the material-jetting lowering path used in the Shore-hardness cross-compilation result. Figure~\ref{fig:results_cross_shore_hardness_j750_model_fit} shows the fitted relationship; full fitting details, coefficients, and error statistics are provided in the supplementary information.

\subsection{Sampled Volumetric Attributes}
\label{sec:results_sampled_volumetric_attributes}

Sampled volumes can serve as source attributes in the same lowering and compilation workflow as analytic fields. Volumetric datasets often carry spatial information on native grids, including simulation fields, graphics volumes, and CT scan data. We import these data as sampled attributes, interpolate them over the object domain, translate them to target attributes, and compile the resolved fields at the resolution required by the selected fabrication toolchain.

\subsubsection{Volumes and OpenVDB}

\begin{figure}[t]
    \centering
    \includegraphics[width=\linewidth]{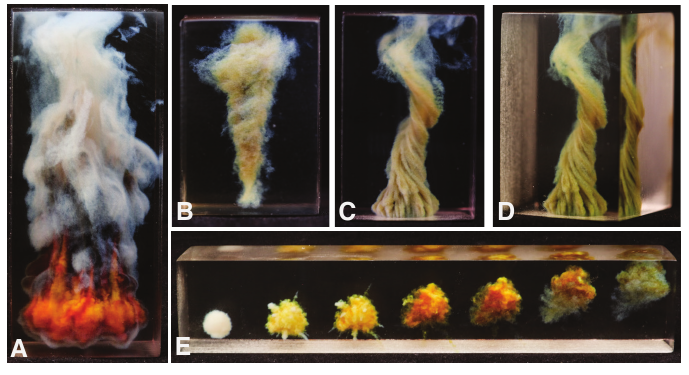}
    \caption{
    Fabrication of OpenVDB volumes as sampled source attributes.
    (A) Fire volume printed from density and temperature fields mapped to RGBA color.
    (B--D) Tornado and dust-devil volumes printed from density fields mapped to color and opacity.
    (E) Explosion sequence printed from density and temperature fields.
    Each sampled volume is compiled using a stochastic color inkjet compiler, demonstrating that simulation-style volumetric data can be resampled into printer-resolution material assignments.
    }
    \label{fig:results_vdb_volumes}
\end{figure}

Figure~\ref{fig:results_vdb_volumes} demonstrates this workflow with OpenVDB volumes from simulation-style datasets~\cite{openvdb_examples, jangafx_free_vdb_volumes}. Scalar density fields, and paired density--temperature fields, are mapped to RGBA color attributes and compiled using the stochastic color inkjet compiler. These examples show that volumetric data commonly used in graphics and simulation can become fabrication source attributes without first converting the volume into segmented regions or surface meshes. This example highlights the method's ability to treat sampled volumetric fields as first-class design attributes for fabrication.

\subsubsection{CT Scans}

\begin{figure}[t]
    \centering
    \includegraphics[width=\linewidth]{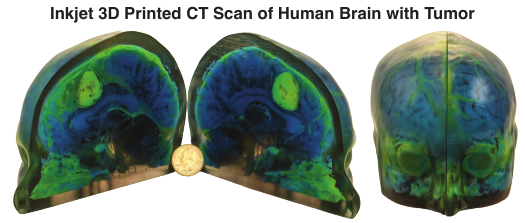}
    \caption{
    Inkjet 3D print of a human brain CT volume with a visible tumor.
    The CT data is imported as a sampled radiodensity attribute and mapped to RGBA color for stochastic color inkjet compilation.
    The printed result preserves continuous radiodensity gradients through the volume rather than reducing the scan to segmented surface meshes.
    }
    \label{fig:results_brain_scan}
\end{figure}

\begin{figure}[t]
    \centering
    \includegraphics[width=\linewidth]{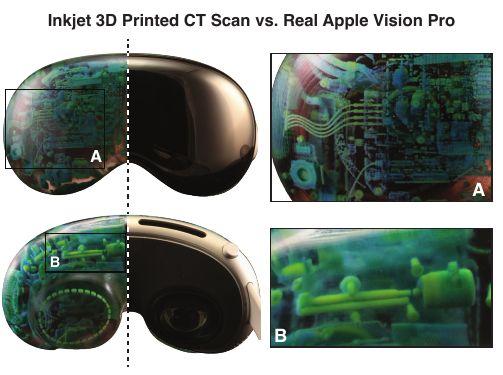}
    \caption{
    Volumetric inkjet print of an Apple Vision Pro CT scan compared with the physical artifact.
    The normalized radiodensity volume is mapped to color and opacity attributes, allowing internal electronic and optical structures to remain visible in the printed model.
    Inset (A) highlights internal wires, circuit boards, and a camera. Inset (B) shows a motor and lead screw used to adjust the interpupillary distance of the displays. An \href{https://matterassembly.org/OpenVCAD-Docs/v3/gallery.html}{interactive side-by-side comparison} is available in the online project gallery.
    }
    \label{fig:results_apple_vision_pro}
\end{figure}

Figures~\ref{fig:results_brain_scan} and~\ref{fig:results_apple_vision_pro} apply the sampled-attribute workflow to CT data from medical and industrial sources. In Fig.~\ref{fig:results_brain_scan}, a brain CT volume is imported as a sampled radiodensity source attribute and mapped to RGBA color for color inkjet fabrication, preserving continuous variation through the tumor, surrounding tissue, and anatomical gradients. In Fig.~\ref{fig:results_apple_vision_pro}, radiodensity from a CT scan of an Apple Vision Pro is mapped to color and opacity, exposing internal wires, circuit boards, cameras, and mechanical components as volumetric visual features. In both cases, the compiler operates directly on continuous scan-derived attributes rather than requiring the data to be reduced to segmented regions, labeled surfaces, or exterior geometry. These examples highlight the method's ability to treat CT volumes as fabrication-ready attribute sources across different scan domains.

\begin{figure}[!b]
\centering
\includegraphics[width=\linewidth]{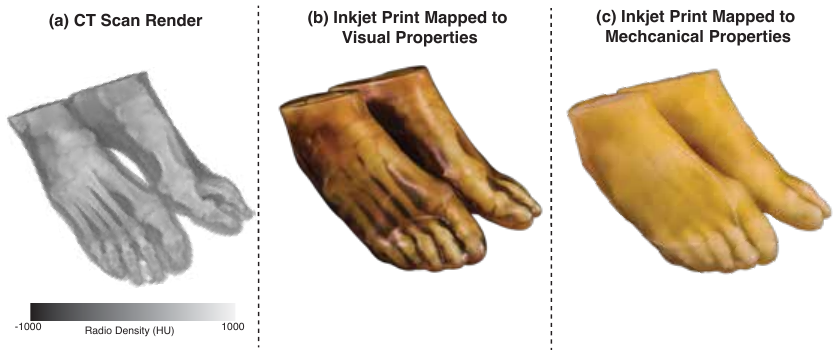}
    \caption{
    CT-derived visual and mechanical attribute fabrication from the same paired-feet scan.
    (a) Source CT volume imported as a sampled radiodensity attribute.
    (b) Inkjet print mapped to visual properties through RGBA color.
    (c) Inkjet print mapped to mechanical properties, including modulus and toughness, which are then resolved to material volume fractions.
    The example demonstrates multi-stage translation from sampled volumetric data to distinct physical outputs. An \href{https://matterassembly.org/OpenVCAD-Docs/v3/gallery.html}{interactive side-by-side comparison} is available in the online project gallery.
    }
    \label{fig:results_feet}
\end{figure}

The previous CT examples compile sampled radiodensity to visual attributes for color inkjet printing. In contrast, Figure~\ref{fig:results_feet} demonstrates how one sampled CT source can drive multiple fabrication targets. Starting from a single CT scan of a pair of feet, we generate two physical realizations. The first maps radiodensity to RGBA color for visual inkjet printing. The second maps radiodensity to mechanical attributes, modulus and toughness, using anatomical-region property targets consistent with the ranges reported by Smith et al.~\cite{smith_digital_2024} and HU-to-anatomy mappings~\cite{schneider_correlation_2000}. These mechanical attributes are then lowered to volume fractions using Smith's model and compiled for material jetting. The resulting print is not only a visual anatomical model, but a multi-stiffness artifact whose mechanical response varies through the volume: the bones inside the toes are rigid, while the surrounding skin and heel regions remain soft and compliant. The full radiodensity$\rightarrow$RGBA color and radiodensity$\rightarrow$modulus+toughness mappings are provided in the supplementary information. This example highlights the compiler's ability to translate one sampled source attribute into distinct visual and mechanical fabrication outcomes.

\subsection{Cross-Compilation Across Fabrication Toolchains}
\label{sec:results_cross_compilation}
Cross-compilation demonstrates an advantage of separating design intent from fabrication-specific realization. We evaluate whether the same source attributes can be realized through different physical mechanisms, compiler requirements, and output formats. This separation is valuable not only across fundamentally different fabrication processes, but also across variants of the same process. For example, material-jetting systems may differ in available materials, voxel encodings, calibration relationships, supported mixing strategies, and machine-facing file formats, particularly across manufacturers and printer generations. By retaining attributes at the source level, these differences can be handled through target-specific translation models and compilers rather than by reformulating the design for each machine.

First, we cross-compile Shore hardness fields to inkjet, where hardness is realized by Vero/Agilus30 material fractions, and to material extrusion, where hardness is realized by foaming TPU process control. Second, we cross-compile an RGB color field to material-jetting and filament-based color workflows. These examples test whether source designs can remain expressed in intended spatial attributes while translation models and compilers resolve the target-specific realization details.

\subsubsection{Shore-Hardness Cross-Compilation}
\label{sec:results_cross_compile_shore_hardness}

\begin{figure*}[t]
    \centering
    \includegraphics[width=\linewidth]{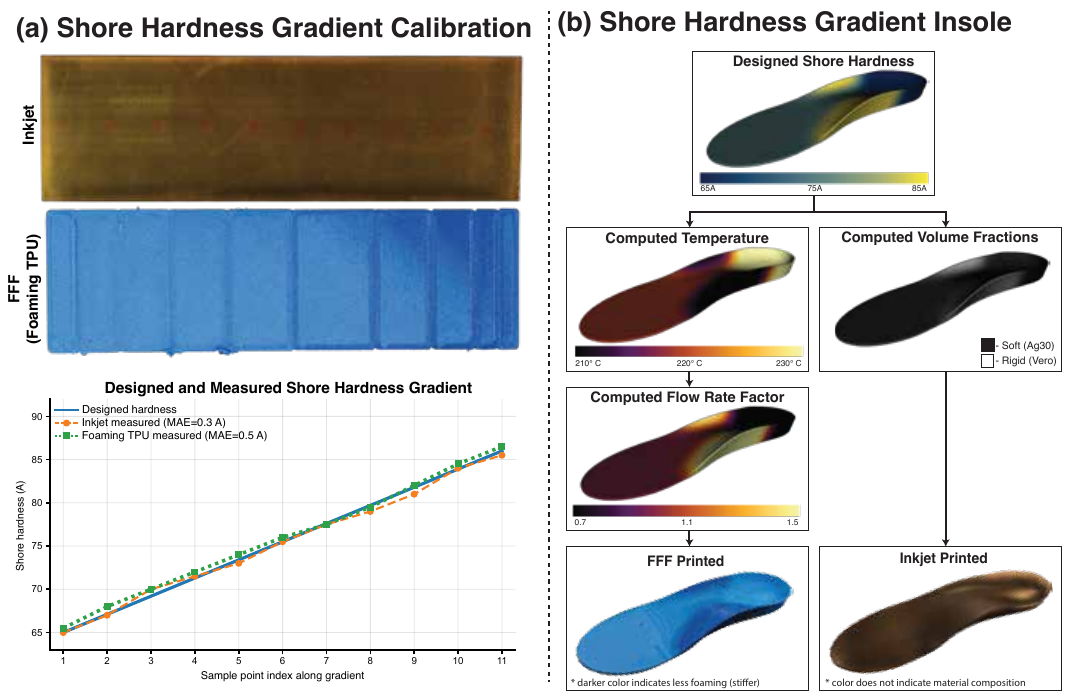}
    \caption{
    Shore-hardness cross-compilation.
    (a) Linear 65A--85A calibration bar fabricated by inkjet and foaming TPU material extrusion, with measured profiles compared against the requested hardness field. Foaming TPU results are reported by Wade et al.~\cite{wade_slicer_project_control} and are provided for comparison.
    (b) Application-scale insole compiled from one shore hardness field to workflow-specific realization attributes for foaming TPU material extrusion and inkjet.
    }
    \label{fig:results_cross_shore_hardness}
\end{figure*}

Figure~\ref{fig:results_cross_shore_hardness} evaluates Shore-hardness cross-compilation from a single high-level attribute specification to two fabrication workflows. The source design for the calibration specimen was specified once as a linear shore hardness field ranging from 65A to 85A over a \(100~\mathrm{mm} \times 30~\mathrm{mm} \times 10~\mathrm{mm}\) bar. The gradient start and end positions were offset \(12~\mathrm{mm}\) from the specimen edges along the \(x\)-axis, allowing measurements to be taken within the geometric constraints of the Shore A protocol. Eleven locations were sampled at regular intervals along the gradient.

For inkjet, the translation graph lowered shore hardness to volume fractions, and the stochastic inkjet compiler converted the resulting Vero/Agilus30 material fractions into discrete voxel assignments. For material extrusion, the same shore hardness field was lowered using the foaming TPU models of Wade et al.~\cite{wade_implicit_2025-1,wade_slicer_project_control} to compute temperature and compensatory flow-rate fields.

Both fabricated bars reproduced the designed hardness profile closely. The plotted values in Fig.~\ref{fig:results_cross_shore_hardness}a are the average measurements at each location. Mean absolute error against the requested hardness profile was 0.3A for inkjet and 0.5A for foaming TPU material extrusion. This agreement shows that the same design-intent field can be resolved into different realization attributes while preserving the requested functional gradient.

Figure~\ref{fig:results_cross_shore_hardness}b applies the same workflow to a full-scale insole. The source design specifies a spatial shore hardness field with a compliant 65A heel region, stiffer 85A arch and lateral-midfoot support regions, and an intermediate 75A region elsewhere, with smooth transitions between zones. From this single source field, the system computes the workflow-specific attributes needed for each compiler: temperature and flow rate for foaming TPU material extrusion, and soft/rigid material fractions for inkjet. Together, the calibration bar and insole demonstrate cross-compilation from one functional specification to distinct fabrication processes, separating modeling intent from translation and compiler-specific realization.

\subsubsection{Color Cross-Compilation}
\label{sec:cross-compile-color}

\begin{figure*}[th!]
    \centering
    \includegraphics[width=\linewidth]{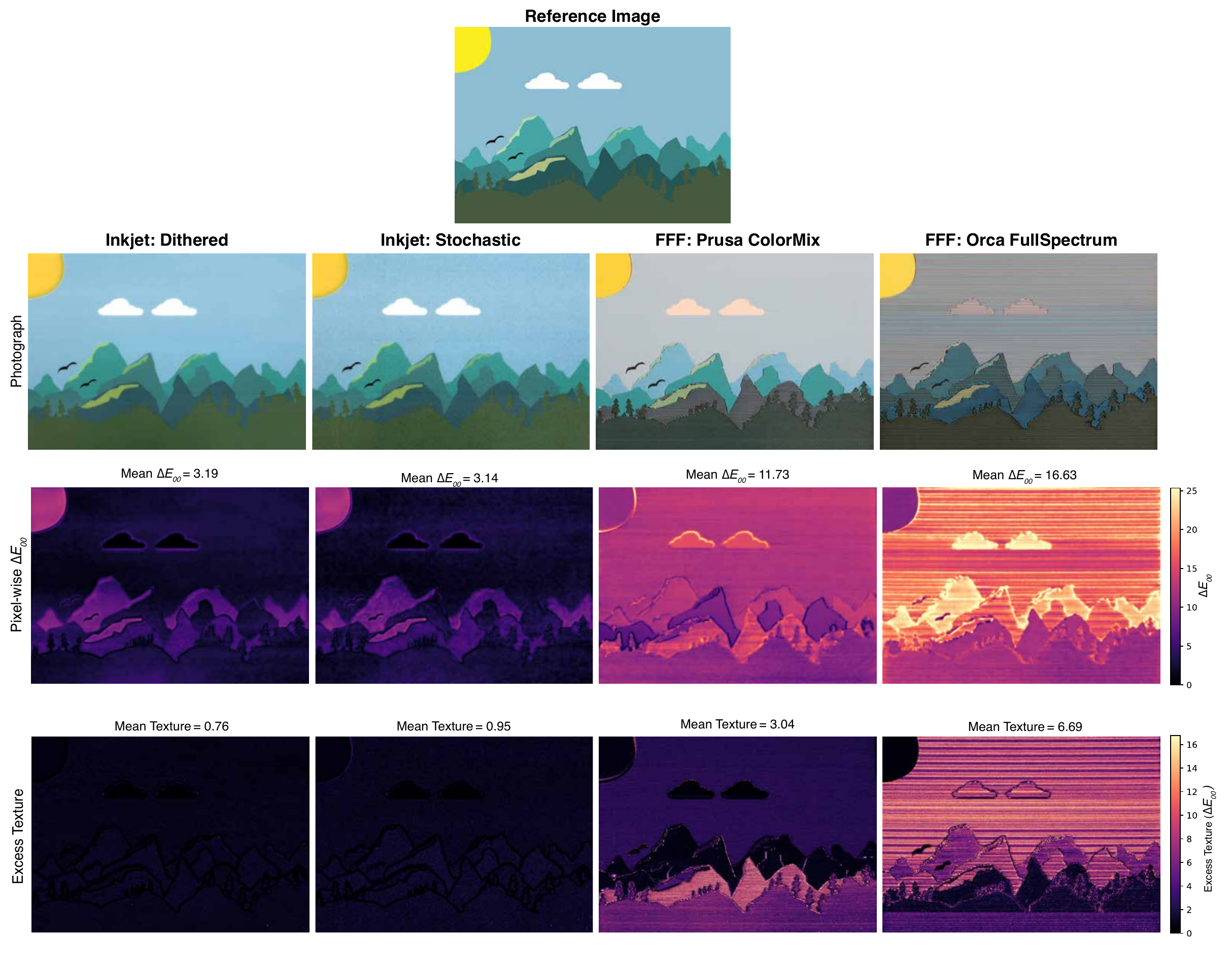}
    \caption{
    Color error and texture residuals for the mountain scene compiled across material-jetting and FFF workflows.
    Row three reports $\Delta E_{00}$ after Gaussian filtering in CIELAB space, emphasizing regional color accuracy.
    Row four reports excess texture residuals, emphasizing print-induced high-frequency variation.
    The material-jetting workflows produced the lowest regional color errors and texture residuals, while the FFF workflows introduced larger color deviations and stronger layer- or halftone-induced texture.
    Mean values are reported in each panel.
    The FFF: Prusa ColorMix and FFF: Orca FullSpectrum photographs, error maps, texture maps, and quantitative measurements were originally reported by Wade et al.~\cite{wade_slicer_project_control} and are reproduced here as comparison points for evaluating cross-compilation across fabrication modalities.
    }
    \label{fig:mountain-perceptual-quality}
\end{figure*}

Figure~\ref{fig:mountain-perceptual-quality} evaluates color cross-compilation using a mountain-scene image as a compact benchmark for comparing visual reproduction across fabrication toolchains. The rendered source design was imported as an RGB color attribute and extruded into a planar 3D object. The same source design was compiled across two printing modalities and four color reproduction workflows. For material jetting, we evaluated structured error-diffusion dithering and stochastic compiler output. For fused filament fabrication on a five-toolhead system, we used the Prusa ColorMix and OrcaSlicer FullSpectrum results reported by Wade et al.~\cite{wade_slicer_project_control}. We reproduce these FFF conditions here as comparison points for evaluating cross-compilation across fabrication modalities.

We quantified color fidelity by photographing each sample under fixed camera settings, lighting conditions, and alignment. Each photograph was registered to the digital reference and compared in CIELAB space using CIEDE2000 color difference, $\Delta E_{00}$. Before computing $\Delta E_{00}$, both the reference and sample images were Gaussian filtered in CIELAB space with $\sigma = 12$ pixels. This metric therefore emphasizes regional color reproduction rather than droplet-, bead-, or pixel-scale variation. We separately measured print-induced high-frequency artifacts using an excess texture residual. This residual isolates high-frequency variation after removing the low-frequency color component, then subtracts the corresponding residual of the reference image so that the reported value emphasizes texture introduced by fabrication rather than texture already present in the source image.

The error maps in Fig.~\ref{fig:mountain-perceptual-quality} show clear workflow-dependent tradeoffs. Inkjet produced the lowest regional color errors, with mean $\Delta E_{00}=3.19$ for structured dithering and $\Delta E_{00}=3.14$ for stochastic output. These workflows therefore achieved similar regional color accuracy. However, structured dithering produced lower excess texture than stochastic output, with mean texture residuals of 0.76 and 0.95, respectively. This indicates that deterministic error diffusion better suppresses local grain in this example, while stochastic compilation provides a simpler and lower-cost computational alternative evaluated further in Section~\ref{sec:results_computational_cost}.

The MEX outputs reported by Wade et al.~\cite{wade_slicer_project_control} produced larger color errors and higher texture residuals than the material-jetting outputs, reflecting the lower spatial resolution, bead-based deposition, and more limited local color control of the evaluated filament workflows. Prusa ColorMix achieved a mean $\Delta E_{00}$ of 11.73 and a mean texture residual of 3.04, while OrcaSlicer FullSpectrum produced a mean $\Delta E_{00}$ of 16.63 and a mean texture residual of 6.69. We include these previously reported measurements to expose process-dependent tradeoffs rather than to present them as independent experimental replications. Material jetting provided higher regional color fidelity and lower texture for this benchmark, while FFF provided a lower-cost and more accessible route to full-color fabrication through slicer-based filament halftoning.

Across the Shore-hardness and color demonstrations, cross-compilation preserves attribute-level intent while exposing workflow-dependent realization quality. The hardness examples show that functional intent can be realized across different material mechanisms with small measured errors. The color examples show that visual intent can be retargeted across different color reproduction workflows, while preserving measurable differences in fidelity and texture. Together, these results demonstrate that source designs can be portable across fabrication toolchains without implying that all workflows realize the design with identical accuracy, resolution, or artifacts.

\subsection{Compilation Cost}
\label{sec:results_computational_cost}

We measured export time to assess whether resolved attribute fields can be compiled at fabrication-relevant scales. Each benchmark reports the wall-clock time required to take a resolved design and export it to the corresponding output format. Resolver time was negligible relative to export time, which is dominated by sampling the resolved geometry and attributes into the target representation, quantizing or discretizing fields as required by the compiler, and writing the output files. Benchmarks were run as single executions on an Apple M5 MacBook Pro using 10 CPU threads.

Table~\ref{tab:results_compilation_cost} reports bounding-box size, total sampled points, export time, and sample throughput for the examples used in the results. The benchmarks consist of the objects from the results section. The demonstrated exports range from \(2.17e7\) to \(3.66e10\) samples. Stochastic inkjet compilation scales to tens of billions of samples for large CT-derived prints, including the paired-feet and Apple Vision Pro examples. Smaller OpenVDB volumes export in approximately one to thirteen minutes. The paired-feet mechanical print is marginally slower than the paired-feet color print at the same sample count because the mechanical workflow evaluates a more expensive translation path before material assignment.

\begin{table*}[t]
    \centering
    \caption{
    Export-time benchmarks for the demonstrated examples. \emph{Samples} denotes the number of spatial evaluations used by the compiler, independent of whether the target output is a voxel stack, process-annotated toolpath, or slicer-project representation. Throughput is reported in millions of samples per second. These benchmarks primarily characterize whether the compiler supports practical fabrication workflows; in the demonstrated cases, export time is dominated by the number of spatial samples evaluated rather than by the geometric size of the object alone.
    }
    \label{tab:results_compilation_cost}
    \scriptsize
    \setlength{\tabcolsep}{4pt}
    \renewcommand{\arraystretch}{1.08}
    \resizebox{\textwidth}{!}{%
    \begin{tabular}{@{}llllrr@{}}
        \toprule
        Example & Compiler & Bounding box (mm) & Samples & Export (min) & Samples/s ($10^6$) \\
        \midrule
        Fire, Fig.~\ref{fig:results_vdb_volumes}A
        & Inkjet--Stochastic
        & $100\times42\times43$
        & 1.86e9
        & 11.4
        & 2.72 \\

        Tornado, Fig.~\ref{fig:results_vdb_volumes}B
        & Inkjet--Stochastic
        & $49\times33\times34$
        & 5.79e8
        & 2.0
        & 4.83 \\

        Dust devil, Fig.~\ref{fig:results_vdb_volumes}C
        & Inkjet--Stochastic
        & $58\times38\times38$
        & 8.57e8
        & 2.6
        & 5.56 \\

        Explosion, Fig.~\ref{fig:results_vdb_volumes}E
        & Inkjet--Stochastic
        & $157\times30\times25$
        & 1.20e9
        & 13.3
        & 1.51 \\

        Brain CT, Fig.~\ref{fig:results_brain_scan}
        & Inkjet--Stochastic
        & $154\times111\times110$
        & 1.95e10
        & 28.20
        & 11.50 \\

        Apple Vision Pro CT, Fig.~\ref{fig:results_apple_vision_pro}
        & Inkjet--Stochastic
        & $183\times95\times178$
        & 3.20e10
        & 22.2
        & 24.00 \\

        Feet CT, color, Fig.~\ref{fig:results_feet}B
        & Inkjet--Stochastic
        & $159\times131\times169$
        & 3.66e10
        & 45.6
        & 13.40 \\

        Feet CT, mechanical, Fig.~\ref{fig:results_feet}C
        & Inkjet--Stochastic
        & $159\times131\times169$
        & 3.66e10
        & 49.4
        & 12.30 \\

        Insole, inkjet, Fig.~\ref{fig:results_cross_shore_hardness}
        & Inkjet--Stochastic
        & $289\times104\times30$
        & 9.44e9
        & 16.6
        & 9.49 \\

        Insole, MEX, Fig.~\ref{fig:results_cross_shore_hardness}
        & MEX--Gradient-informed
        & $290\times105\times31$
        & 1.19e8
        & 1.2
        & 1.72 \\

        Mountains, Fig.~\ref{fig:mountain-perceptual-quality}B
        & Inkjet--Stochastic
        & $149\times106\times10$
        & 1.64e9
        & 3.3
        & 8.32 \\

        Mountains, Fig.~\ref{fig:mountain-perceptual-quality}C
        & Inkjet--Dithered
        & $149\times106\times10$
        & 1.64e9
        & 5.6
        & 4.87 \\

        Mountains, Fig.~\ref{fig:mountain-perceptual-quality}D
        & MEX--ColorMix
        & $149\times106\times10$
        & 2.17e7
        & 17.9
        & 0.02 \\

        Mountains, Fig.~\ref{fig:mountain-perceptual-quality}E
        & MEX--FullSpectrum
        & $149\times106\times10$
        & 2.17e7
        & 17.4
        & 0.02 \\
        \bottomrule
    \end{tabular}%
    }
\end{table*}

The mountain-scene benchmark isolates the cost of compiler choice because the inkjet stochastic and dithered outputs use the same object size and sample count. Stochastic inkjet export required 3.3 min for \(1.64e9\) samples, while dithered inkjet export required 5.6 min for the same sample count. This difference reflects the tradeoff observed in the color-quality results: structured dithering reduced local texture artifacts, but its error-diffusion procedure introduced additional export cost. The MEX color workflows sampled far fewer points because they operate at slicer-project resolution rather than inkjet voxel resolution. Their throughput values are therefore not directly comparable to inkjet throughput; these exports are dominated by mesh construction and \texttt{.3MF} project generation rather than dense voxel sampling.

These measurements are provided only as general references for computational scale, not as direct performance comparisons. Earlier programmable multi-material fabrication systems addressed different problem formulations and did not support the cross-compilation of a common attribute-based source design across multiple fabrication backends. OpenFab reported full-volume synthesis for 0.12B--14.2B voxels at 300 DPI, including a 1.8B-voxel butterfly in 33 min and a 1.2B-voxel marble table in 25 min~\cite{vidimce_openfab_2013}. The closest inkjet-scale examples in Table~\ref{tab:results_compilation_cost}, including the 1.64B-sample mountain object and the 1.20B-sample explosion volume, fall within a similar sample-count regime. The larger CT-derived examples extend to tens of billions of samples. Although differences in representation, implementation, hardware, and supported workflows preclude direct benchmarking, these references indicate that the proposed workflow operates at fabrication-relevant scales while additionally supporting typed attributes, sampled source data, modular translation models, and multiple compiler targets.

\section{Conclusion}

Heterogeneous additive manufacturing needs representations that preserve design intent without forcing designers to author printer-specific material recipes, process parameters, or slicer settings directly. This paper establishes typed attribute fields as a design-intent layer for heterogeneous fabrication. Geometry is represented as an implicit object domain, while visual, measured, mechanical, material, process, and slicer-level quantities are represented as named, typed spatial fields over that domain. This separation lets designers specify what an object should express, while translation models and compiler modules determine how that intent is realized for a selected fabrication toolchain.

The central contribution is a staged workflow for heterogeneous design and fabrication. Typed attributes provide a common spatial interface for heterogeneous information. Composition rules define how these fields behave under hierarchy, transforms, Boolean operations, and overlapping regions. Translation models lower source or intent attributes into realization attributes when valid calibrated, empirical, analytical, or user-defined mappings exist. Compiler modules then request the attributes they require and sample the resolved fields into process-specific outputs, including voxel material stacks, process-annotated G-code, and slicer-project files. Together, these components separate source design, attribute realization, and machine-specific output.

The results show the value of this separation. Sampled CT and OpenVDB volumes enter the same workflow as analytic fields and can be translated into visual or mechanical attributes for fabrication. We also demonstrate cross-compilation: one shore hardness field is realized through both inkjet material fractions and foaming-TPU process control, and one RGB color field is realized through both material-jetting and filament-based color workflows. These examples show that the source design can remain stable while the realization attributes, compiler outputs, and fabrication mechanisms change. They also show that portability does not erase process-specific behavior; accuracy, texture, resolution, cost, and fabrication artifacts remain measurable consequences of the selected workflow.

Our results and examples reflect available models and compilers, but the method is extensible to additional model and compiler types. New applications require material data, process calibrations, analytical mappings, or learned models that relate intent attributes to realizable material or process fields. They also require compiler modules for additional fabrication processes. Future work should therefore expand translation-model libraries, add compiler targets beyond the inkjet and material-extrusion workflows demonstrated here, and accelerate high-resolution compilation with GPU implementations. These extensions would make the approach applicable to broader domains, including pre-surgical planning models, soft robotics, functionally graded materials, and multifunctional structures.

We provide the representation, translation framework, compiler interface, and demonstrated examples as an open-source Python package implementation. The goal is to make the system extensible: researchers can add attribute types, sampled-data sources, calibrated mappings, and fabrication compilers as new materials and processes become available. By organizing heterogeneous fabrication around typed spatial intent rather than fixed material recipes, this work provides a foundation for reusable design specifications, measurable process-specific realization, and reproducible research in multi-material and functionally graded additive manufacturing.

\section{Acknowledgments}
This material is based upon work supported by the Charles Stark Draper Laboratory, Inc. under Contract No. N00030-24-C-6001. Any opinions, findings and conclusions or recommendations expressed in this material are those of the author(s) and do not necessarily reflect the views of Strategic Systems Programs.

\appendix
\section{Appendix}
All examples presented in this work, together with the Python and C++ implementations, are available at \url{https://matterassembly.org/openvcad}. OpenVCAD supports Windows, macOS, and Linux and can be installed using \texttt{pip install OpenVCAD}.

\bibliographystyle{plain}
\bibliography{references,other_sources} 

\end{document}